\def\papertitle{Cybercasing 2.0: You Get What You Pay For}
\def\paperkeywords{}

\documentclass[twocolumn,10pt]{article}
\usepackage{makeidx}
\usepackage[latin1]{inputenc}
\usepackage[pdftex]{graphicx}  
\usepackage{lastpage}
\usepackage{multirow}
\usepackage{pifont}
\usepackage{subfigure}
\usepackage{times}
\usepackage{xspace}
\usepackage{color}
\usepackage[usenames,dvipsnames,svgnames,table]{xcolor}
\usepackage{url}

\definecolor{linkcol}{rgb}{0,0,1}
\definecolor{citecol}{rgb}{0,0.5,0}
\definecolor{urlcol}{rgb}{0.3,0,0}

\newcommand{\note}[1]{{}}
\renewcommand{\note}[1]{{\textcolor{red}{[\textit{#1}]}}}

\newcommand{\imc}[1]{}\xspace

\usepackage[bookmarks=true,%
bookmarksnumbered=true,%
colorlinks=true,%
linkcolor=linkcol,%
citecolor=citecol,%
urlcolor=urlcol,%
hypertexnames=true,%
pdfpagelabels]{hyperref}

\hypersetup{
pdfauthor = {Choi et al.},
pdftitle = {\papertitle},
pdfkeywords = {\paperkeywords},
pdfcreator = {LaTeX with hyperref package},
pdfproducer = {pdflatex}}

\title{\papertitle}

\author{Jaeyoung Choi\and Istemi Ekin Akkus\thanks{Work done while visiting ICSI.} \and Serge Egelman\and Gerald Friedland\and Robin Sommer\and Michael Carl Tschantz\and Nicholas Weaver \vspace{0.25ex}\and
International Computer Science Institute}

\begin{document}

\maketitle

\begin{abstract} 

Under U.S. law, marketing databases exist under almost no legal
restrictions concerning accuracy, access, or confidentiality.  We
explore the possible (mis)use of these databases in a criminal context
by conducting two experiments.  First, we show how this data can be
used for ``cybercasing'' by using this data to resolve the physical
addresses of individuals who are likely to be on vacation.  Second, we
evaluate the utility of a ``bride to be'' mailing list augmented with
data obtained by searching both Facebook and a bridal registry
aggregator.  We conclude that marketing data is not necessarily
harmless and can represent a fruitful target for criminal misuse.

\end{abstract}

\section{Introduction}
\label{sec:intro}

In the United States, numerous laws have been enacted to protect very specific classes of information about individuals: 
the Fair Credit Reporting Act (FCRA) protects information used to determine credit worthiness~\cite{fcra:ftc}; %
the Health Insurance Portability and Accountability Act (HIPAA) protects medical records~\cite{hipaa03summary}; 
the Family Educational Rights and Privacy Act (FERPA) protects education records~\cite{ferpa:ed}.
However, the U.S.\ lacks comprehensive privacy protections.  In particular, databases used for marketing purposes lack accountability or oversight.

Numerous marketing databases and marketing data brokers
 provide information with little or no restrictions, and
sometimes even on a free-trial basis.  The database records often contain names, addresses, email addresses, interests, and numerous other pieces of information about an individual. 
These databases are often created by the bulk purchasing of commercial transaction records and/or aggregating public records~\cite{ftc14report}.  Considered in the abstract and by themselves, most individuals are unlikely to have expectations of privacy about these data sources.  However, the aggregation and cross-referencing of this data from multiple sources is likely to raise privacy concerns.

The piecemeal privacy regulations in the U.S. currently do not cover data used for this purpose.  Indeed, from a business standpoint, the biggest constraint on this
data is that it \emph{must not} be used for purposes that would bring
it under the FCRA (or other regulations).  That is, the brokers purposefully avoid regulations that might bring accountability.  
At the same time, these brokers are aware of how unsettling their practices are to consumers: while purchasing this type of data, we observed that some data brokers insist on source confidentiality.  That is, if a consumer inquires, the recipient must not disclose the identity of the data broker from which the data was purchased.

Despite these databases being beyond the reach of privacy regulations in the U.S., such as the FCRA, and doing their best to remain as such, the information they collect is potentially dangerous.  
We investigate some possible criminal uses of these marketing
databases, especially when combined with information obtained from web
searches.  We studied two possible attacks.  First, we utilize email to
address mappings in conjunction with YouTube searches. Next, we purchased a list of ``brides to be'' from a data broker and correlated this information with publicly available data on Facebook.
 Both experiments reproduce previous
``cybercasing''~\cite{Friedland2010,Friedland2011} experiments, which examined how geotags could be abused.  Unlike geotagging,
however, a person has effectively no control over the release of data
by marketing data brokers.
We hope our study will encourage regulators to look at regulating such marketing databases more closely.

\section{Prior Work}

Friedland and Sommer previously performed several studies to examine how publicly available data could be misused~\cite{Friedland2010}.  Their studies focused on high-accuracy location information attached as meta-data to audio, image, and video files. Specifically, they examined how geotags could be used for ``cybercasing,'' using online data and services to mount real-world attacks.

The first scenario involved tracking a specific person, in this case TV show
host Adam Savage who is an active Twitter user.
It turned out that most images posted to his feed contained an exact geolocation
attached by his smartphone, allowing them to locate his studio, places where he
walks his dog, his home, and also where he met with other celebrities 
while on travel. 

In the second scenario, the authors inspected a random sample of Craigslist postings containing geotagged images. They examined all postings to the San Francisco Bay Area's \emph{For Sale} section over a period of four days,
in total collecting 68,729 images, of which about 1.3\,$\%$ of the images were tagged
with GPS coordinates.  A fair amount of the geotagged postings offered
high-valued goods, such as diamonds apparently photographed at home, making
them potential targets for burglars. In addition, many posters even offered
specifics about when and how they wanted to be contacted (``please call Sunday
after 3pm''), which allowed for speculation about when that person might or might not be at
home.  

In a third scenario, the authors examined whether one can semi-automatically identify
the home addresses of people who normally live in a certain area but are
currently on vacation.  Such knowledge offers opportunities for
burglars to break into their unoccupied houses. They wrote a script using the
YouTube API that, given a home location, a radius, and a keyword, finds a set
of matching videos shot within this radius and containing the keyword.
For all the videos found, the script then gathers the
associated YouTube user names and downloads all of their videos that are a certain 
\emph{vacation distance} away but have been uploaded within the last couple of days.
The home location was set to be in downtown Berkeley, CA, and the radius to 100\,km. 
The authors searched for the keyword ``kids'' since many people publish home
videos of their children. The vacation distance was 1000\,miles. Even though
only about 3\,$\%$ of the YouTube content was geotagged at the time, the script reported 1000
hits (the maximum number the site returns for any query) for the initial set
of matching videos. These then expanded to about 50,000 total videos in the
second step identifying all other videos from the corresponding users. 106 of
these turned out to have been taken more than 1000 miles away and uploaded the
same week. Sifting quickly through the titles of these videos, the authors easily found that about a dozen looked promising for a successful burglary.

Friedland and Choi built on this work by removing the need for geotags using an automatic location estimation system~\cite{SC2011}.  Their approach to location estimation was a machine-learning and semantic-web driven method based on the open service \url{GeoNames.org}. GeoNames covers all countries and contains 8 million entries of place names and corresponding geo-coordinates. It provides a web-based search engine and an API which returns a list of matching entries ordered by their relevance to the query.  They showed how geotags are unnecessary for cybercasing by searching for videos that contained keywords of known cities, and then correlating any names found in the videos with phone book data.

Friedland et al.\ examined these methods more generally by showing how multiple data sources could be aggregated to make better inferences~\cite{Friedland2011}.  In this manner, they showed how criminals could use multiple public data sources to increase the likelihood that a potential burglary target will not be home, improve a stalker's reach, or even to frame someone.

While prior work in this area examined combining data from multiple sources to increase a criminal's effectiveness, no one has yet explored how data brokers might fit into this ecosystem.  Unlike prior work, in which more privacy awareness would have been beneficial for users (e.g., disable geotagging), there is not much users can do for the release of consumer data by data brokers~\cite{ftc14report}. We posit that the market for consumer data creates a potential boon for criminals well beyond 
the control of consumers and what has previously been discussed in the literature.

\section{The Ecology of Data}

\textbf{The Light Side:} Whenever a user interacts with a company, such
as buying a product or filling out a sweepstakes form, that creates a
data footprint.  A product order tells the company the person's name, shipping address, billing address,
email address, and what they purchased.  The company might then sell this data
element to data brokers without the customer's knowledge or explicit permission.

These data brokers coalesce, analyze, filter, aggregate, and resell
the resulting data, with each broker attempting to create a more accurate
profile of all individuals in their data set.  
While data can sometimes end up corrupted, especially for those with
``common-name@big-provider'' email addresses that others might
mistakenly use, it can often provide accurate information about some individuals.  
Some estimates suggest there are 4000 separate companies
involved in this process~\cite{databroker_estimate}, and many brokers make the data available to any
buyer willing to pay.

Access to a broker's data usually occurs in one of two forms, an append
interface or provided lists~\cite{ftc14report}.  An append interface has the buyer
providing a list of records and the data broker annotates the results
with other features if available, charging for each successful
annotation.  
For example, the data customer might provide a
list of email addresses, to which the broker will append features such as
demographic information, purchase habits, income estimates, home
ownership, or other fields.  In particular, some brokers specifically
support annotating mailing addresses, allowing the data customer to associate
email addresses to mailing addresses.

Some companies also offer interactive access for appending
information.  Rapleaf provides an API-based interface where a customer
can provide either email addresses or mailing addresses and receive a
demographic profile in a claimed 50ms of processing time~\cite{Rapleaf}.  Demographic
parameters include gender, age range, income level, home ownership,
and various interests such as sports, travel, pets, outdoor and
adventure, and whether the person tends to donate to charitable
causes.

The second form of access is to provide lists about individuals matching a given
criteria, such as ``brides to be'' or ``rape victims''
(in a notorious case, which
the provider subsequently insisted was simply a test, not an actual list~\cite{hill13forbes}%
), often with additional constraints such as zip code or domain
specific data, such as wedding date.  The data broker then provides
an agreed upon number of matching entries and the specified fields,
such as email address and mailing address.

\textbf{The Dark Side:} Criminals have discovered the benefits of
aggregating and reselling identifying and financial data
 for the purposes of identity
theft.  Credit report data is remarkably cheap, with a full target
credit report costing a reported \$15 giving the target's full name,
address, date of birth, and social security
number (SSN)~\cite{krebs:ssndob:cost}.
The service also offers to provide someone's SSN and date of birth given their name and
address for just \$1.50.

There was also a report back in 2007 of criminals using marketing
lists to find elderly scam victims for telemarketing
fraud~\cite{criminal_use}.  While marketing lists are certainly not a new technology, we have recently reached a point where ubiquitous online data can be augmented with these lists to create unprecedented views into every aspects of an individual's life.  In an attempt to draw attention to this issue, we performed two experiments using the information that we purchased from data brokers.

\section{Study One: How's The Trip?}

We initially set out to determine whether we could reproduce our
previous cybercasing~\cite{Friedland2010} experiment without utilizing
geotags.  In the previous study, we searched for vacation videos
with geotags, and then discovered home videos from the same account
with geotags.

In our new study, we began by searching for videos based on
the vacation topic (list) and extracting the Google username.  After
excluding obviously bogus usernames, we submitted 2824 names to a data
broker as an append request.
(We do not name the data broker we used since our contract with it appears to prohibit disclosing its identity.)
The overall cost was a \$500 setup fee plus an additional \$0.10 for each match successfully
appended.

The result was surprisingly negative: out of the 2824 addresses
submitted, only 9 were successfully appended.  We believe this is due
to three factors: a lack of correlation with purchasing behavior, the
list-focused nature of the data broker we utilized, and the relative
quality of this particular data broker.

First, if a user doesn't utilize their Google account
for making purchases, there will be no link between the mailing
address and email address available to a data broker to sell.  Google
itself may have information about the user's address, but Google has
no incentive to sell this information to others as it represents a
competitive advantage.

Second, list-focused data brokers do not prioritize complete coverage as highly as more traditional data brokers, such as credit reporting agencies.
In credit reporting, a small number of brokers strive for complete
coverage.
If a credit
agency only had information on 50\% of the population, it would not be
competitive in the marketplace since the data consumers select the queried names.  A list-centric data broker's
incentives are different: they don't need complete coverage, rather
they need quality in the data they have since the broker gets to select its best data matching the requested criteria to share with the data consumer.
This broker in particular focuses as a reseller of lists with a wide
variety of topics, including religious affiliation (Catholic, Jewish,
Islamic, etc), economic profile (credit score), ethnicity, political
donation habits, holders of handgun concealed carry permits, and even
such esoteric lists as ``boat owners in LaGuna Niguel, California.''

Third, we selected this broker mostly due to setup cost.  Most data
brokers are only interested in large orders.  Even this broker
required a \$500 setup cost for the append query, and this may
represent a case of ``you get what you pay for.''

We also did some spot checking on results, and found that append data
may be of marginal quality.  For example, although it correctly
identified one author's father's address and one cousin, the address
for another cousin and the author himself were completely wrong: not
even in the correct state.

\section{Study Two:  What a Happy Bride}

List purchases, however, don't suffer the same defects: not only do
the providers claim high accuracy (often over 90\%, and sometimes as
high as 95\%), but the nature of list construction prevents the ``null
entry'' problem faced when purchasing append data.  Thus, we considered
the possibility of lists as targets for theft.  Numerous criteria,
ranging from known gun owners to any selection criteria for high
income might have potential.

For our study, we chose ``brides to be,'' with name, mailing
address, email address, and date of wedding.  The cost of this data
was only \$0.20 an entry for 5000 entries.  Having the email address allows
some additional searching for ancillary data and indicates an
online presence, the mailing address gives the person's location, and
the wedding date itself provides a day when the person's home
will likely be empty.  We explore this ancillary data as a means of
estimating list accuracy.

Our first check was to determine whether we could find bridal
registries for the listed names, as a lower bound on the correctness.
We utilized \url{registry.weddingchannel.com}, a bridal registry aggregator
service that indexes multiple bridal registries and allows search by
name, with the returned information including city, state, and date of
wedding.

We uses several matching criteria.  A strong match, where full name,
city, state, and wedding date matched, and a weaker match where the
full name, city, and state match but the wedding date does not.
One quarter of the names featured a strong match, with an additional 7.4\%
obtaining a weaker match.

Of particular note, however, is that the pair of first name and last name did
not match at all in 41\% of the cases.  Given the breadth of the
wedding registry aggregator itself (16 separate registries), this
suggests that either there are other items feeding into the bridal list (i.e., beyond companies selling their bridal registries to data brokers) or that
the bridal list has an error rate significantly higher than that claimed by
the data broker.

We also checked if some registries seems overrepresented in the data.
The most significant matches were with Bed Bath \& Beyond (942), Target
(643), and Macy's (406).  
We could not draw any conclusions about the broker's sources given the overall popularity of these stores and the lack of domination by any one of them.

Facebook is also a rich source of ancillary data.  Previously,
Facebook's Graph API enabled searching by email, but this interface is now
deprecated.  Instead, we searched on a combination of first name, last name, and city, which
is a fuzzier match. First, we retrieved a list of user matches using Facebook Graph API with queries '[first name] [last name] [city name]' with 'user' as the search type. Then for each user in the retrieved list, we parsed each of the user's timeline HTML to gather the current city and state information, since these pieces of information are not available when only using the Graph API. We were able to obtain candidate Facebook
accounts for 17.1\% of the list entries that had a first name, last name, city and state match.

We then manually examined a portion of the Facebook matches.  
We examined 129 data-list entries that also had a strong match on the bridal registry list.
Amongst them, we found 64 (50\%) to have clear indications of an upcoming wedding.
We also examined 290 data-list entries that lacked a matching registry entry.
We found 107 (37\%) to have such indications.

Of particular note, 50 Facebook pages included photographs of the
bride's wedding or engagement ring, an indicator of income level,
while 5 pages included information about the bride's honeymoon plans.

\section{The Rise of Criminal Brokers?}

Criminal brokers already exist for finical data acting as ``append''
services, but there is nothing stopping similar services from
developing for the sorts of non-financial data we examined.  Given the lower barriers to access,
it would be straightforward for criminal groups to set up their own
 data brokers.

The likeliest target would be criminal lists, akin to the marketing
lists, of high candidate potential victims.  For burglary or similar
activity, a subscription service could provide lists by zip-code of
possible candidates with associated profiles.  

The lists themselves don't need to be too expensive to be profitable.
Since our purchases cost~\$0.20 a name, if only 1/50th of the names
are salable, the selling cost of such lists needs to be only \$10/name
for the criminal broker to break even.  Likewise, the consumer of the
list doesn't need to obtain much more than \$10 of value from a name for
a \$10 purchase to be worthwhile.

The best target is probably gun ownership.  Within the criminal black
market, guns represent a unique product, where there isn't a
substantial loss in value when attempting to fence.  Although a
gun-ownership list doesn't provide a set of times when a target is
away, it does provide a list of homes which contain particularly
valuable items.

Obtaining the lists we used was straightforward: a legitimate
looking email address (we used our own \url{.edu} address when purchasing
data) and a credit card to buy the data.  The biggest obstacle was
learning the correct terms when communicating with the data brokers.

Overall, the biggest limitation is list accuracy.  We have trouble
believing the 90\% accuracy rate claimed by the list brokers, but its
also clear from the bridal data that this does achieve an accuracy rate that appears to be roughly 50\%
accuracy.  List inaccuracy increases the overall cost to the attacker,
as any false positive in the list represents wasted resources when an
attacker evaluates the result.

For marketeers, inaccuracy is a minor but tolerable cost: a false
match is a wasted mailer, but the total cost per mismatch is only a
dollar or so.  Criminal uses may have a higher penalty for mismatch:
if someone needs to investigate a target in person, a false match
might have a cost measured in tens or perhaps even hundreds of
dollars.

We've shown that, to at least some degree, list inaccuracy is
countered with ancillatory data.  For our bridal example, we used
Facebook or registry services to validate portions the raw data.  The
ancillatory data, especially if it has a single sided error (people
seldom post about a nonexistent wedding on Facebook), is of particular
use since it acts to ensure a sub-list of true positives.  Low cost
validation strategies may depend on the context but, when available,
can produce a much cleaner data stream.

One of the most powerful tools that we did not investigate is Google
Maps streetview.  The lists already contain the target's address,
which makes a simple matter of putting the address into Google to
instantly gauge any obvious security systems (such as signs), the
income level, and secondary signals such as the bumper stickers of
cars in the driveway (as it is highly unlikely for a truck with an NRA
bumper sticker to be owned by a non-gun owner).

\section{Conclusion}
\label{sec:conclusions}

We live in a soup of data, producing little eddies of information with
every action we take.  A whole host of data brokers exist to slurp up
this information, process it, and digest it into a form enjoyed by
marketeers and merchants.

Yet this data, although not generally regulated by the U.S.\ government,
is not without its risk.  We showed the ability to partially replicate
the previous cybercasing result without requiring any geotagged data,
an exercise that will probably grow in precision as marketeers attempt
to map email to physical location on a more regular basis.

We also showed the possibility of creating criminal lists derived from
a public purchased list of brides to be, and how such lists can be
both enhanced and cleaned using ancillatory search data such as
Facebook profiles and bridal registry information.  Overall, we
believe that marketing data is not necessarily harmless:
there is significant potential for abuse.

\section*{Acknowledgements}
This research was supported by the U.S.\ National Science Foundation
(NSF) grants CNS 1065240 and CNS 1514509.  The views and conclusions
contained in this document are those of the authors and should not be
interpreted as representing the official policies, either expressed or
implied, of any sponsoring institution, the U.S.\ government or any
other entity.

\bibliographystyle{plain}
\bibliography{paper}

\begin{thebibliography}{10}

\bibitem{krebs:ssndob:cost}
{Brian Krebs}.
\newblock {Credit Reports Sold for Cheap in the Underweb}.
\newblock
  \url{http://krebsonsecurity.com/2013/03/credit-reports-sold-for-cheap-in-the-underweb/}.

\bibitem{criminal_use}
{Charles Duhigg}.
\newblock { Bilking the Elderly, With a Corporate Assist }.
\newblock
  \url{http://www.nytimes.com/2007/05/20/business/20tele.html?_r=1&pagewanted=print}.

\bibitem{ftc14report}
{Federal Trade Commission}.
\newblock Data brokers: A call for transparency and accountability, 2014.

\bibitem{SC2011}
Gerald Friedland and Jaeyoung Choi.
\newblock Semantic computing and privacy: A case study using inferred
  geo-location.
\newblock {\em International Journal of Semantic Computing}, 5(01):79--93,
  2011.

\bibitem{Friedland2011}
Gerald Friedland, Gregor Maier, Robin Sommer, and Nicholas Weaver.
\newblock {S}herlock {H}olmes' evil twin: On the impact of global inference for
  online privacy.
\newblock In {\em Proceedings of the 2011 New Security Paradigms Workshop},
  NSPW '11, pages 105--114, New York, NY, USA, 2011. ACM.

\bibitem{Friedland2010}
Gerald Friedland and Robin Sommer.
\newblock Cybercasing the joint: On the privacy implications of geo-tagging.
\newblock In {\em Proceedings of the 5th USENIX Conference on Hot Topics in
  Security}, HotSec'10, pages 1--8, Berkeley, CA, USA, 2010. USENIX
  Association.

\bibitem{hill13forbes}
Kashmir Hill.
\newblock Data broker was selling lists of rape victims, alcoholics, and
  `erectile dysfunction sufferers'.
\newblock Forbes, December 19, 2013.
\newblock
  \url{http://www.forbes.com/sites/kashmirhill/2013/12/19/data-broker-was-selling-lists-of-rape-alcoholism-and-erectile-dysfunction-sufferers/}.

\bibitem{hipaa03summary}
{Office for Civil Rights}.
\newblock Summary of the {HIPAA} privacy rule.
\newblock OCR Privacy Brief, U.S.\ Department of Health and Human Services,
  2003.

\bibitem{Rapleaf}
Rapleaf.
\newblock {Rapleaf API}, 2013.
\newblock \url{https://www.rapleaf.com/developers/personalization-api/}.

\bibitem{ferpa:ed}
{U.S. Department of Education}.
\newblock Family educational rights and privacy act ({FERPA}).
\newblock 20 U.S.C. \S 1232g.
\newblock \url{http://www.ed.gov/policy/gen/guid/fpco/ferpa/index.html}.

\bibitem{fcra:ftc}
{U.S. Federal Trade Commission}.
\newblock Fair credit reporting act.
\newblock 15 U.S.C. \S 1681 et seq., September 2012.
\newblock
  \url{http://www.consumer.ftc.gov/sites/default/files/articles/pdf/pdf-0111-fair-credit-reporting-act.pdf}.

\bibitem{databroker_estimate}
{Yasha Levine}.
\newblock {What Surveillance Valley knows about you}.
\newblock \url{http://pando.com/2013/12/22/a-peek-into-surveillance-valley/}.

\end{thebibliography}
\end{document}